\documentclass[preprint2]{aastex}

\usepackage[dvips]{color}

\usepackage{longtable}

\shorttitle{X-ray Substructures in Clusters of Galaxies}
\shortauthors{Andrade-Santos et al.}

\begin{document}


\title{A New Method to Quantify X-ray Substructures in \\ Clusters of Galaxies}



\author{Felipe Andrade-Santos, 
Gast\~{a}o B. Lima Neto, Tatiana Ferraz
Lagan\'{a}}
\affil{Universidade de S\~{a}o Paulo, Departamento de Astronomia, Instituto de
\\ Astronomia, Geof\'{i}sica e Ci\^{e}ncias Atmosf\'{e}ricas, Rua do Mat\~{a}o 1226,
\\ Cidade Universit\'{a}ria, 05508-090, S\~{a}o Paulo, SP, Brazil}







\begin{abstract}
We present a new method to quantify substructures in clusters of galaxies,
based on the analysis of the intensity of structures.
This analysis is done in a residual image that is the result of the
subtraction of a surface brightness model, obtained by fitting a
two-dimensional analytical model ($\beta$-model or S\'{e}rsic profile) with
elliptical symmetry, from the X-ray image.
Our method is applied to 34 clusters observed by the \textit{Chandra} Space
Telescope that are in the redshift range $z
\in[0.02,0.2]$ and have a signal-to-noise ratio greater than 100.
We present the calibration of the method and 
the relations between the substructure level with
physical quantities, such as the mass, X-ray
luminosity, temperature, and cluster redshift. 
We use our method to separate the clusters in two sub-samples of high and low
substructure levels. 
We conclude, using Monte Carlo simulations,  that the method  recuperates
very well the true amount of substructure for small angular core radii clusters 
(with respect to the whole image size) and good signal-to-noise observations.
We find no evidence of correlation between the substructure level and physical
properties of the clusters such as mass, gas temperature, X-ray luminosity
and redshift. The scaling relations for the two sub-samples (high and low substructure
level clusters) are different (they present an off-set, i.e., given a fixed mass or temperature,
low substructure clusters tend to be more X-ray luminous), which is an important result for cosmological 
tests using the mass-luminosity relation to obtain the cluster mass function, since 
they rely on the assumption that clusters do not present different scaling 
relations according to their dynamical state.
\end{abstract}



\section{Introduction}

Clusters of galaxies are the largest virialized objects in the Universe, the
upper limit of collapsed halo mass function.
In an Universe dominated by a cosmological constant and cold dark matter
($\Lambda$CDM), dark matter halos are formed by gravitational instability from
primordial quantum fluctuations in the mass density field. The amplitude of
those fluctuations increases as they cease expanding with the Hubble flux,
collapse and virialize, forming dense and relaxed structures. Smaller
structures grow to larger ones through mergers, up to clusters of
galaxies in the present time. In this hierarchical scenario of structures
formation, clusters are thus dynamically young objects and contain evidence of
their recent past merging history \citep[e.g.][]{1993Kauffmann}.
We can relate substructures with the cluster dynamical age
\citep[e.g.][]{1992Richstone,2003Suwa}: the more substructure (their total intensity) a cluster
presents, the younger (dynamically speaking) it is.


The hot intra-cluster plasma is a powerful X-ray source and its observation
reveals the projected spatial distribution of most of the baryonic mass. X-ray
studies of galaxy clusters are thus particularly relevant in this context, as
they can give us clues to the dynamical age of clusters (e.g.,
\citet{2000Henriksen} -- Abell 3266, \citet{2003LimaNeto} --
Abell 970, \citet{2005Ferrari} -- Abell 3921). Analysis of
substructure in the intra-cluster plasma spatial distribution should help us
determining the dynamical state of galaxy clusters. A very good review about
the theory and observational status of the study of substructures based on
X-ray data in clusters of galaxies is given by \citet{2005Jeltema}. Here, we
only briefly discuss some of the previous work on cluster substructures.

\citet{1992Jones} made the first X-ray systematic study of structures in
galaxy clusters, visually analyzing 208 objects observed by the
\textit{Einstein} satellite, establishing that merging must be a common
phenomenon in clusters. \citet{1992Richstone} developed in an original
theoretical study a relation between substructures and cosmology, where they
put constraints on cosmological parameters by the fractional rate of major
mergers in clusters.

X-ray surface brightness allows us to perform statistical tests such as
centroid and ellipticity variation \citep{1995Mohr}, relating the
dynamical age of clusters with its morphology. \citet{1995Buote,1996Buote}
developed a method to quantify X-ray substructures in clusters of galaxies
from the moments of the expansion in Fourier series of the X-ray surface
brightness. \citet{2005Jeltema} used the same method, referred to as the
\textit{power-ratio method}, in a sample of 40 clusters of galaxies observed
by \textit{Chandra}. They showed that clusters in general are less relaxed at
$z>0.5$, than at $z\simeq 0$.

Semi-analytic methods give an indication of the expected evolution of cluster
substructure and its dependence on cosmological parameters, however, the best
method of constraining cosmological models is probably through the comparison
with hydrodynamic cluster simulations. For instance, \citet{2003Suwa} compared
simulated clusters in a $\Lambda$CDM and an OCDM cosmology, at both $z = 0$
and $z = 0.5$, using several methods for quantifying structure. They restrict
themselves to comparing the ability of different statistical indicators in
distinguishing different simulated cosmologies, showing that cluster structure
can potentially constrain $\Omega_\Lambda$ or the dark energy equation of
state.

Although a lot of effort has been done in order to advance our understanding
about substructure in clusters of galaxies, from theoretical to numerical
simulation studies, we propose in the present work
a novel method of quantifying substructures that has a simple physical
interpretation: the substructure level, the way it is defined, reflects the
fraction of the total X-ray luminosity that is emitted by the substructures,
 serving
as a tool to understand the underlying physical processes taking place
during the cluster evolution.


This paper is organized as follows. In section 2, we describe the sample
selection and in section 3 the data reduction and analysis are discussed. In
section 4, the substructure level is defined and our method is described, with 
its calibrations being discussed in section 5. In
section 6 the results are presented and discussed and conclusions are finally
presented in section 7. The cosmology assumed in this paper is given by
$\Omega_M=0.3$, $\Omega_{\Lambda}=0.7$ and $H_0=70$~km~s$^{-1}$Mpc$^{-1}$.


\section{Sample Definition}

\begin{figure}[!htb]
\centering 
\scalebox{0.38}{\includegraphics{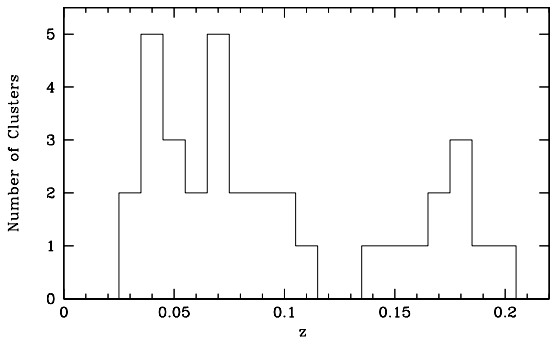}}
\caption[]{\small Histogram of the redshift distribution of clusters in our sample.}\label{hist_z}
\end{figure}

Our method was applied to 34 clusters observed by the \textit{Chandra} X-ray
Telescope ACIS-I detector, with signal-to-noise ratios ($S/N$) greater than 100,
and that are in the redshift range $z \in[0.02,0.2]$. 
Figure \ref{hist_z} shows the cluster redshift
distribution. There is an apparent gap within $z \in[0.12,0.14]$. This is
due to the incompleteness of the sample, but there is no particular
redshift interval with an excess of objects and our results do not
depend on the sample completeness.

The criteria for clusters selection were chosen so as to ensure a suitable
signal-to-noise ratio and a large enough image to work upon, without
introducing bias for specific clusters. However, biases that we do not control
may affect our sample. Clusters are observed in time-competitive telescopes,
so they must present something ``special'', many times
substructures and irregularities, that make them ``worth'' being observed.
Therefore one should keep in mind this caveat, that it is possible that our
sample may have a tendency to present more substructures than the average
expected for all clusters in the redshift range $z \in[0.02,0.2]$.

\section{Data Reduction and Analysis}

In order to obtain calibrated images without artifacts, adequate to 
be used with our method of substructure detection, 
it is necessary to follow a series of procedures of cleaning and
filtering the X-ray data. Otherwise, we would have contamination that would be
detected as spurious substructures.


\subsection{Data Reduction} 

We have used the package CIAO 3.4. Initially a level 2 events file has been
generated from a level 1 events file, using the standard pipeline
procedure\footnote{http://cxc.harvard.edu/ciao3.4/threads/createL2/} and the
calibration files, CALDB 3.3.0. Periods with high particle background
(flares) were excluded using the \textit{lc$\_$clean} script.
%
%
At this point, a re-binned image with pixels corresponding to 16 raw physical
pixels (4x4, which roughly corresponds to 2'' pixels) is created from the new level 2 event file, in the energy band
0.3 to 7.0~keV.
Then, we produce exposure maps and use them to obtain flat images from which
the source points are removed by filling circles around each source with
a random Poisson sampling with the same distribution as found in a
 circular region
close to the source. Finally,
we fit a 2D analytical surface brightness model.


\subsection{Surface Brightness}

The surface brightness profile is the projection of the plasma emissivity
along the line of sight. We will assume two radial analytical profiles for the
surface brightness: the $\beta$-model \citep{1976Cavaliere} and the S\'{e}rsic
\citep{1997Pislar,2003Demarco}.

In order to take into account the ellipticity of the plasma emission we use
the following standard coordinates transformation:
\begin{displaymath}
\label{BS_coord}
\left\{ \begin{array}{l}
x' = (x-x_0) \cos\theta - (y-y_0) \sin\theta;\\
y' = (x-x_0) \sin\theta + (y-y_0) \cos\theta;\\
r^2 = x'^2 + \displaystyle \frac{y'^2}{(1-\epsilon)^2},\\
\end{array} \right.
\end{displaymath}
where $(x_0,y_0)$ is the X-ray emission center coordinates, $\theta$ is the
position angle, and $\epsilon$ is the ellipticity.

The $\beta$-model may now be defined as follows:
\begin{displaymath}
\label{BS_beta}
\Sigma(r)= \Sigma_{0} \left[ 1 + 
       \left(\frac{r}{r_{c}}\right)^{2} \right]^{-3\beta+0.5} + b,
\end{displaymath}
where $r_{c}$ is the core radius, $\beta$ is the shape parameter, and
$\Sigma_{0}$ is the central surface brightness. The parameter $b$
corresponds to the background, and is supposed to be constant throughout the image
(hence the importance of the exposure map correction).

The S\'ersic model is defined as follows:
\begin{eqnarray}
\label{BS_sersic}
\Sigma(r)= \Sigma_{0} \exp \left[-\left(\frac{r}{a}\right)^{\nu}
 \right] + b, \nonumber
\end{eqnarray}
where $a$ is the scale parameter, $\nu$ (often represented as $1/n$) is the
shape parameter and $b$ is again the background surface brightness.

Once we have the image correctly processed, we fit a 2D surface brightness
model to it using a standard minimum squares method, $\chi^2$, and obtain the residual
image, which is going to be the starting point for substructure
quantification.

We fitted the $\beta$ and S\'ersic models for most of the clusters, and in the
case where both models were fitted, we chose to use the one that gave the
smaller substructure level (see below how the substructure level is defined
and computed). In practical terms, this is the same as choosing the fit with
the smallest $\chi^2$.
%
%
The 2D surface brightness model fitted for each cluster is
presented in Table~\ref{clusters_parameters}.


\section{X-ray Substructures}

Previous studies on ICM substructure have been
done, either qualitatively \citep{1984Jones,1992Jones,
2008Lagana,2010Lagana} or
quantitatively \citep{1992Richstone,1995Buote,2005Jeltema}, based on different
techniques. There is, however, no method that takes into account the ratio
between the number of counts on the residual and on the original images, which
will be referred to as the \textit{residual flux method}. We
describe here this method to quantify the substructure 
on the intra-cluster plasma emission.
%


\subsection{Substructure Level}

We start by defining a threshold for the residual image in order
to identify the pixels
which had a number of counts statistically significant above or below
(positive and negative residues) the 2D surface brightness fitted model at the
pixel position. The threshold in each pixel was defined as the square root of
the number of counts of the model in the correspondent pixel, i.e., the 
expected variance.

Then we quantify the substructure level by
computing the ratio between the total number of counts of the residual and original images (taking
the absolute value of the negative counts in the residual image and treating them exactly
as the counts in the positive regions - after selecting those (in absolute value) above the threshold). By construction,
the substructure level, $S$, is defined as:
\begin{eqnarray}
S\equiv \frac{\displaystyle\sum_{i=1}^n \left|C_i^r\right|}
{\displaystyle\sum_{i=1}^n C_i^t},\label{equation_sub_level}
\end{eqnarray}
where $C_i^r$ is the number of counts of the $i$-th
residual image pixel and $C_i^t$ is the number of counts of the $i$-th
image pixel and $n$ is the number of pixels of the image.

We defined the substructure
level this way because it has 
a direct physical interpretation: it reflects the fraction of the
total X-ray luminosity provided by substructures.

The statistical uncertainties in the substructure level were computed 
using Monte Carlo simulations as described in the section \ref{Monte_Carlo}.





\section{Calibration of the Method}\label{calibration_section}



\subsection{General Case}

We may write equation (\ref{equation_sub_level}) as:
\begin{eqnarray}
S(t)\equiv \frac{\displaystyle\sum_{i=1}^n
\left|C_i^t(t)-M_i^\prime \times t\right|}
{\displaystyle\sum_{i=1}^n (C_i^t - b_i)\times t},\label{equation_sub_level2}
\end{eqnarray}
where $M^\prime$ is the model fitted to the image, which is decomposed
into the cluster surface brigthness model and a constant background,
i.e., $M^\prime = M + b$. 

The number of counts of the \textit{i-th} pixel, for a certain
exposure time $t$, may be written as:
\begin{eqnarray}
C_i^t(t) = P\left[
\left(b_i+
S_i^1+
S_i^2\right)\times t
\right],
\label{cit}
\end{eqnarray}
where $b_i$, $S_i^1$ and $S_i^2$ are the expected number counts in the
\textit{i-th} pixel for an exposure time of $t=1$, in an arbitrary time
unit, from the background, primary cluster and substructures,
respectively. $P(x)$ is the random Poisson deviate of the expected value
$x$. In the limit
when $x \rightarrow t$, $P(x) \rightarrow t \pm t^{1/2}$.

By injecting equation (\ref{cit}) in equation
(\ref{equation_sub_level2}) and taking into account that
the sum of counts of the model is equal to the sum of counts of
the main cluster plus substructures, i.e, $\sum_{i=1}^n M_i =  
\sum_{i=1}^n (S_i^1 + S_i^2)$, we have:
\begin{eqnarray}
S(t)= \frac{\displaystyle\sum_{i=1}^n
\left|P\left[\left(b_i+
S_i^1+
S_i^2\right)\times t \right]
-(M_i+b)\times t\right|}
{\displaystyle\sum_{i=1}^n (S_i^1 + S_i^2) \times t},\label{final_sub_eq}
\end{eqnarray}
where $b$ is the mean background level, i.e., $b=\frac{1}{n}\sum_{i=1}^n b_i$.


\subsection{Long Exposure Time Observation}

We now consider the limit of a very long exposure time.
In this case, $P[(b_i + S_i^1 + S_i^2)\times t] \rightarrow 
(b_i + S_i^1 + S_i^2)\times t$, so the equation (\ref{final_sub_eq}) takes the form:
\begin{eqnarray}
\lim_{t \to \infty} S(t)= \frac{\displaystyle\sum_{i=1}^n
\left|\left(
S_i^1+
S_i^2\right) - M_i - (b - b_i)
\right|}
{\displaystyle\sum_{i=1}^n (S_i^1 + S_i^2)}.\label{final_sub_eq_inf}
\end{eqnarray}

We may write the model as: $M_i = S_i^1 + D_i$, where $D_i$ is the
deviation on the \textit{i-th} pixel due to $S_i^2$ (the presence of substructures
will change the model fitted in the \textit{i-th} pixel by $D_i$). Now the equation 
(\ref{final_sub_eq_inf}) takes the form:
\begin{eqnarray}
\lim_{t \to \infty} S(t)= \frac{\displaystyle\sum_{i=1}^n
\left|
S_i^2 - D_i - (b - b_i) 
\right|}
{\displaystyle\sum_{i=1}^n (S_i^1 + S_i^2)},\label{final_sub_eq_inf2}
\end{eqnarray}
which is different from the ideal case,
\begin{eqnarray}
S^\prime= \frac{\displaystyle\sum_{i=1}^n
\left|
S_i^2 
\right|}
{\displaystyle\sum_{i=1}^n (S_i^1 + S_i^2)},\label{final_sub_eq_inf3}
\end{eqnarray}
in which the substructure level reflects exactly the fraction
of counts provided by the substructures. However, using Monte Carlo simulations 
(which will be discussed in section \ref{Monte_Carlo}) to introduce substructure on 
model images of the clusters of the sample,
one may correct this effect by
introducing a normalization factor in equation (\ref{final_sub_eq_inf2})
for each cluster, allowing us to better estimate the true substructure level 
and quantify the systematic uncertainties involved in this method.    


\subsection{Short Exposure Time Observation}

We consider now
the limit when we have a very short exposure time.
In this case the Poisson noise dominates over the expected value,
 $P(t) \simeq t^{1/2}$, so the numerator of equation 
(\ref{final_sub_eq}) is dominated by noise, that is, it scales 
with $t^{1/2}$, while the denominator scales with $t$, so
$S(t) \propto t^{1/2}/t = t^{-1/2}$. 

\begin{center}
\begin{figure}[!htb]
\centering
\includegraphics[width=0.45\textwidth]{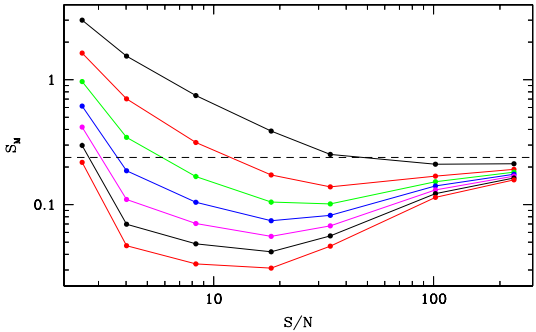}
\caption{\small{Measured Substructure vs. Signal-to-Noise Ratio. All of the curves
correspond to a Gaussian smooth of 1 pixel kernel, and from the top to
the bottom, the thresholds are: 0.5, 0.75, 1, 1.25, 1.5, 1.75,
2 times the square
root of the expected value for each pixel of the model. 
The dashed line corresponds to the actual substruture level, i.e.,
the actual number of counts that is provided by the substructures. 
Colors are used for better
visual distinction between the curves.}\label{SL_vs_S/N}}
\end{figure}
\end{center}

\begin{figure}[!htb]
\centering 
\scalebox{0.38}{\includegraphics{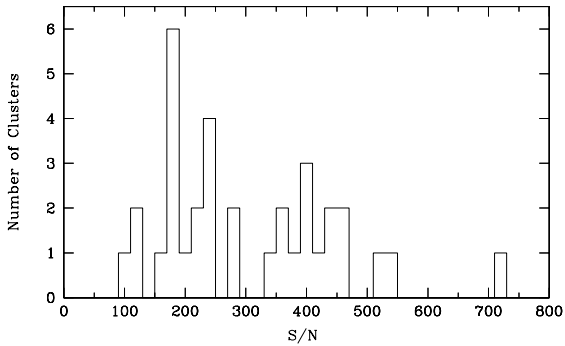}}
\caption[]{\small Signal-to-noise ratio distribution of the clusters.}\label{hist_SNR}
\end{figure}

This property of the substructure level leads to the question: what is the minimum
signal-to-noise ratio required for the method to be applied?
In order to answer this question we must create images of a cluster with substructure and vary the signal-to-noise
to analyse how the substructure quantification varies. With this in mind
we created images of different signal-to-noise of a cluster generated by a $\beta$-model, with $\beta = 2/3$ and 
core radius equal to 20 pixels (in an image of 500 $\times$ 500 pixels). We added
a substructure 15 pixels away from the center 
of the main cluster. It has the same $\beta$ and half the core radius and central surface brightness of the main cluster.  
Figure~\ref{SL_vs_S/N}
shows how the measured substructure level varied with signal-to-noise ratio
for different threshold levels (0.5, 0.75, 1, 1.25, 1.5, 1.75, 2 times the square
root of the expected value for each pixel of the model). The residual images were smoothed 
using a Gaussian kernel of 1 pixel, which corresponds
to $\simeq 2''$. The dashed horizontal 
line corresponds to the actual substruture level, i.e.,
the actual number of counts that is provided by the substructure, and we see that
the measured substructure level converges asymptotically to this value. However, it 
converges differently according to the threshold used, hence the necessity to
calibrate the method to obtain the factor for each cluster (according to the Gaussian
smooth and threshold used) that will correct the value measured.
The value we chose was $(S/N)_{min} \simeq
100$, although the method could also be applied to clusters with worse 
signal-to-noise ratio, which of course would increase the uncertainties.
Figure \ref{hist_SNR} shows the signal-to-noise ratio distribution of the 
sample, which contains only clusters with signal-to-noise ratio greater than 
100. 
 We also want to stress that 
Figure \ref{SL_vs_S/N} depends on the setup of the cluster and substructures and that it is purelly 
illustrative to show the different behavior of the measured substructure level as a function of the threshold
used and the signal-to-noise ratio. 


\subsection{Monte Carlo Simulations}\label{Monte_Carlo}

For each cluster, we
generated an image of the main component using the best fit analytical model. 
Then, we populated the images with substructures having random positions and intensities. A constant background was also added and to all image components were added a white noise following a Poisson distribution. For each cluster, 200 realizations were done.

Once the simulated images were generated, the same procedure used to real 
cluster images was applied for all simulated images. We thus obtained a distribution for the substructure level
 that we compared to the actual level of substructure that was input into the simulated images, which we have control.


The substructures added to the analytical images had surface 
brightnesses described by a $\beta$-model, with core radii and central
surface brightness intensities that could
vary between 25\% to 75\% of the modeled cluster, the exact value being 
determined by a random variable. 
The number of substructures could also vary from 0 (i.e., no substructure) to 3.

\begin{figure*}[!ht]
\centering
\includegraphics[width=1.\textwidth]{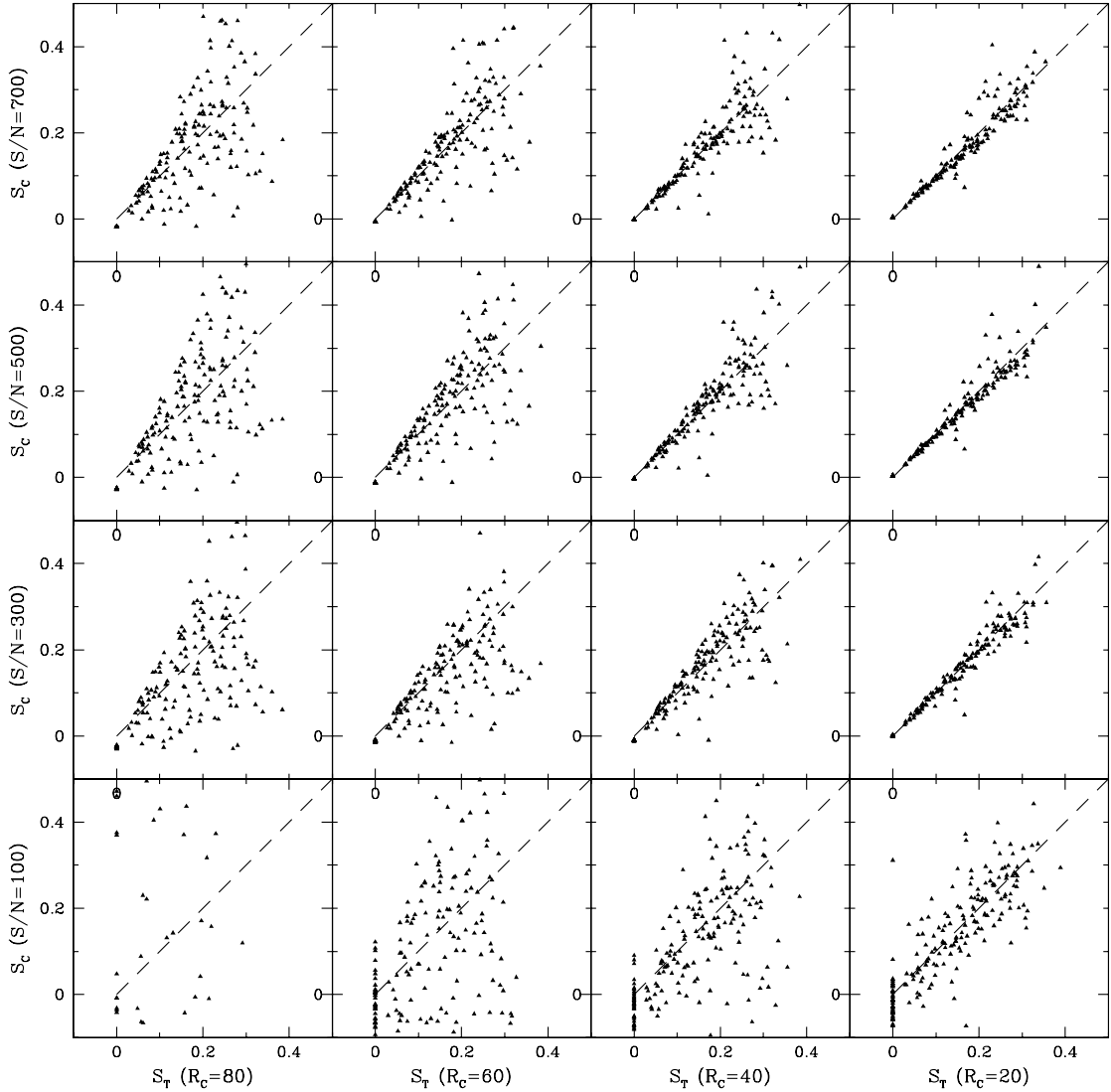}
\caption{\small{Illustration of the sensitivity of the method. For each plot we have the corrected ($S_C$) against the true substructure level ($S_T$) for each of the 200 simulated images. Each plot corresponds to different signal-to-noise ($S/N$) and core radii ($R_c$) simulated clusters, with $R_c$ given in pixels. The simulations were performed in 500 $\times$ 500 pixels images (see Figure \ref{A85_subs_on_model}). For comparison, the dashed lines represent the 1:1 relation between 
the corrected subtructure level against the actual value.}
\label{SN_Rc}}
\end{figure*}

In order to show that basically the quality of the 
substructure quantification depends on the size of the cluster compared to 
the whole image and the signal-to-noise ratio, we present Figure \ref{SN_Rc}
which shows how the corrected substructure level compares to the true
values, for different cluster configurations, in which different synthetic clusters were created, with fixed $\beta = 2/3$, core radius spanning from 20 to 80 pixels (whole image is 500 $\times$ 500 pixels)
and signal-to-noise ratios varying from 100 to 700. We see in the left bottom plots that when
substructures are close to the center of the clusters the method does not give
a reliable result since the substructure is incorporated into the model when 
the surface brigfhtness fit is performed. Therefore, small angular core radii tend to
give better results since the amount of substructure which falls within the 
clustercentric distance is small. 

First, we made a linear fit of the measured substructure level against 
the true substruture level, i.e. for each cluster we had a relation: 
$S_M = a + b \times S_T$, where $S_M$, $S_T$, $a$ and $b$ are the measured and  true 
substructure levels, linear and angular coeficients, respectively.
Once the fit was done, the 
corrected substructure level was computed by:  $S_C = (S_M - a) / b $. In Figure
\ref{3plots_4clusters} we see the measured substructure level plotted against 
the true substructure level on the left panel and then the correction plotted on
the center panel. 

\begin{figure}[!ht]
\centering
\includegraphics[width=0.45\textwidth]{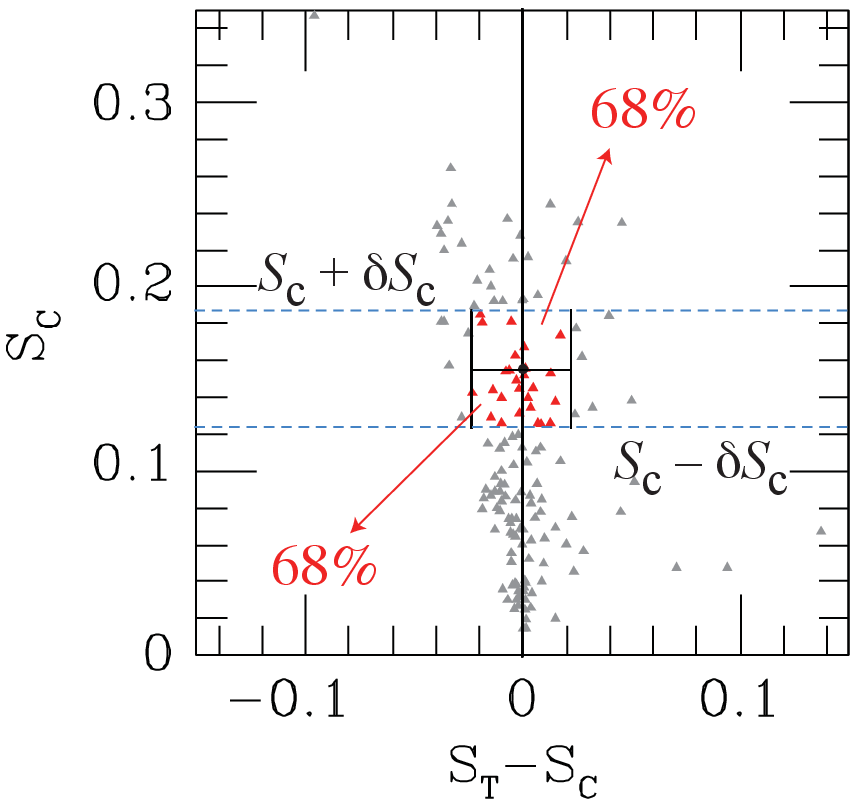}
\caption{\small{Illustration on how the uncertainties on the substructure level are computed.}
\label{error_bar}}
\end{figure}
\begin{figure*}[!ht]
\centering
\includegraphics[width=1.15\textwidth]{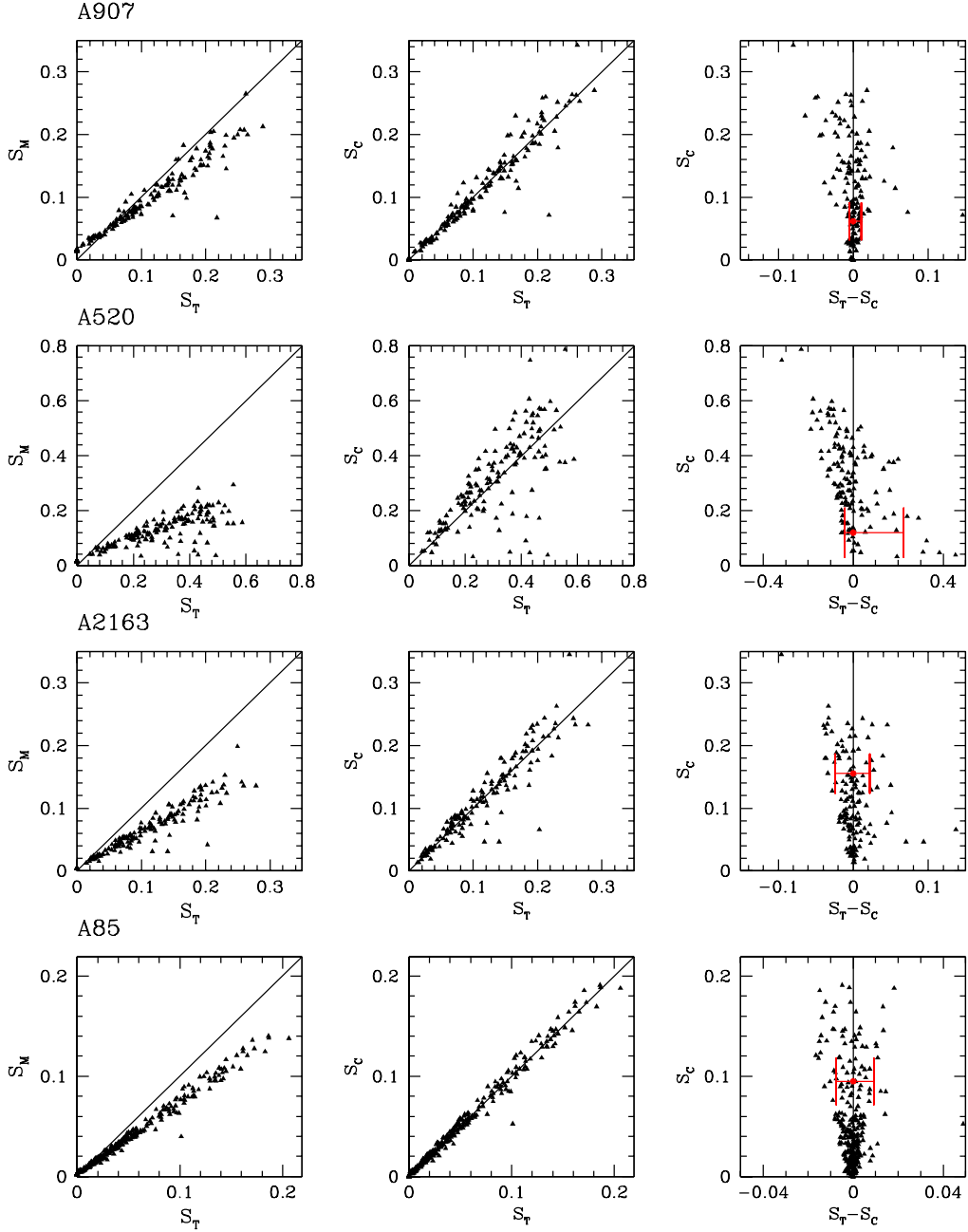}
\caption{\small{Illustration of the calibration applied to some clusters of the sample. \textit{Left:} Measured Substructure,  $S_M$, vs. 
the True Substructure Level, $S_T$, for all 200 Monte Carlo simulations.
\textit{Center:} Corrected Substructure, $S_C$,  vs. 
True Substructure Level for all simulations.
\textit{Right:} Substructure corrected  vs. 
(True Substructure Level $-$ Corrected Substructure ) for the simulations. We only show examples for 4 clusters. 
See text for details on the corrections made. For comparison, the lines represent the 1:1 relation between 
the measured and corrected subtructure level against the actual value.}
\label{3plots_4clusters}}
\end{figure*}

The error bars were determined from the points distribution shown in Fig. \ref{error_bar}.
Starting with the cluster corrected substructure level, $S_C$, we defined a
symmetrical region $S_C \pm \delta S_C$  (horizontal dashed lines in Fig. \ref{error_bar}) where
we have at least 18 data points\footnote{Few points could determine the uncertainties erroneously, whereas too many would
use points corresponding to very different corrected substructure level, so we decided (empirically) to use 18.}
in each side with respect to $S_T-S_C = 0$. Then
the asymmetrical error bars correspond to the range of 68\% of the points in
each side separately (red points in Fig. \ref{error_bar}).

\begin{figure*}[!ht]
\centering
\includegraphics[width=1.00\textwidth]{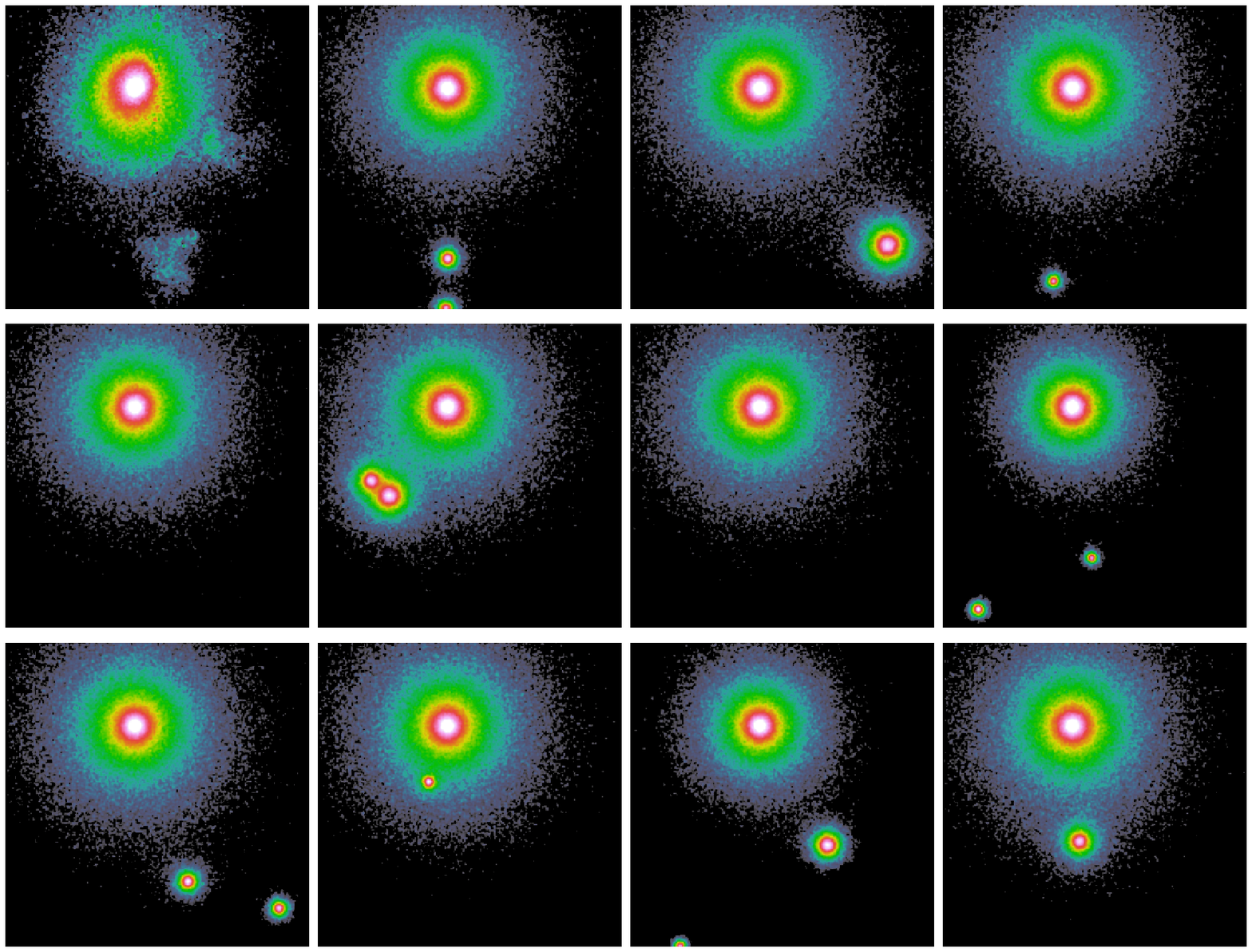}
\caption{\small{On the top left corner the X-ray image of Abell 85 is presented 
along with some of its simulated images containing arbitrary substructure.
}
\label{A85_subs_on_model}}
\end{figure*}

In Figure \ref{3plots_4clusters} we present the calibration for some (four) clusters of the sample, 
which were chosen 
because they represent different levels of substructure (from Abell 907 with $S_C = 0.062$ to Abell 2163 with $S_C = 0.155$) and different core radii and signal-to-noise. This figure shows 
both the measured against the true amount of substructure (left panel)
and the corrected substructure level (central panel) and the method we used for computing the uncertainties
on the substructure quantification (right panel). In Figure \ref{A85_subs_on_model}
we ilustrate the Monte Carlo simulation with a very small sub-sample of the images 
created to calibrate the method for Abell 85. On the top left we see its X-ray image, 
as observed by the \textit{Chandra} Space Telescope, and its simulated images 
containing randomly distributed substructures. 

\section{Results and Discussion}

Now that we have measured and corrected the substructure level and estimated 
the error bars within 68\% confidencel level using Monte Carlo simulations,
we may look for correlations between the substructure level, as 
we defined, and physical properties of the clusters.

For correlations to be correctly assessed, it is necessary to well understand
how variables are related. Linear regression is a fundamental and frequently
used tool in astronomy and it may seem surprising that such a statistical
procedure, apparently simple, may be complicated and controversial \citep[see,
e.g.,][for reviews]{1990Isobe,1992Feigelson,2010Hogg}. Briefly, when the scientific
question clearly asks how one variable depends on the other,
it is more appropriate to use OLS$(Y\vert X)$, Ordinary Least Square - the least square
fit of the function Y(X),
to quantify how the variables are correlated, with Y being the dependent
variable. However, when the scientific question does not clearly identifies
the dependent variable, then it is recommended the used of OLS (Bisector)
which is the bisector between the OLS$(Y\vert X)$ and OLS$(X\vert Y)$ fits,
the last case representing the fit inversion with respect to the variables.

With respect to the size of the sample, when the size is small ($N<50$, $N$
the number of data points), resampling methods such as Jackknife or Bootstrap
should be used \citep{1992Feigelson} to fit the data and estimate the uncertainties. 

The correlation strength is estimated by the Pearson coefficient
\citep[see,][]{1988Pearson}, where its absolute value resides between 0 and 1, with
1 meaning total correlation and 0 none. The interpretation of the correlation 
strength depends on the context. A correlation of 0.9 may be very low if we are 
verifying a physical law with high quality equipments, but may be seen as very
high in social sciences for example, where there are many contributions of complicated
variables. In Table \ref{estatistica} we give the two-tailed null hypothesis significance 
for each Pearson correlation coefficient (See \citet{1992NumRec} for more information on how 
it is computed).  

In our case, we used the OLS$(Y\vert X)$ to fit relations between the substructure level and 
physical parameters, whereas we used the OLS (Bisector) for the scaling relations, 
all fits performed using the Jackknife resampling method. 


\begin{table*}[!ht]
\centering
\begin{tabular}{lllll}
\hline
Relation & Best Fit & Fit & Pearson & Null Hypothesis \\
\hline
$M_{500}$ - $T$ & $M_{500}=0.312^{+0.039}_{-0.036}\times T^{1.64\pm 0.07}$ & OLS (Bisector) & 0.94 & $< 10^{-6}$ \\
$L_X$ - $M_{500}$ & $L_X=0.281^{+0.069}_{-0.055}\times M_{500}^{~~1.94\pm 0.17}$ & OLS (Bisector) & 0.89 & $< 10^{-6}$ \\
$L_X$ - $T$ & $L_X=0.029^{+0.013}_{-0.009}\times T^{3.20\pm 0.29}$ & OLS (Bisector) & 0.91 & $< 10^{-6}$ \\
\hline
\multicolumn{5}{c}{\small{Scale Relations (High Substructure Level)}} \\
\hline
$M_{500}$ - $T$ & $M_{500}=0.378^{+0.079}_{-0.065}\times T^{1.49\pm 0.10}$ & OLS (Bisector) & 0.92 & $< 10^{-6}$ \\
$L_X$ - $M_{500}$ & $L_X=0.424^{+0.145}_{-0.108}\times M_{500}^{~~2.03\pm 0.15}$ & OLS (Bisector) & 0.88 & $< 10^{-6}$ \\
$L_X$ - $T$ & $L_X=0.058^{+0.024}_{-0.017}\times T^{3.02\pm 0.18}$ & OLS (Bisector) & 0.90 & $< 10^{-6}$ \\
\hline
\multicolumn{5}{c}{\small{Scale Relations (Low Substructure Level)}} \\
\hline
$S$ - $(1+z)$ & $S= 0.036^{+0.010}_{-0.007} \times (1+z)^{5.45 \pm 1.95}$ & OLS $\rm Y(X)$ & 0.35 & 0.042439 \\
$S$ - $M_{500}$ & $S= 0.021^{+0.010}_{-0.007} \times M_{500}^{~~0.62 \pm 0.20}$ & OLS $\rm Y(X)$ & 0.49 & 0.003265 \\
$S$ - $T$ & $S= 0.016^{+0.012}_{-0.007} \times T^{0.74 \pm 0.30}$ & OLS $\rm Y(X)$ & 0.36 & 0.036499 \\
$S$ - $L_X$ & $S= 0.040^{+0.014}_{-0.010} \times L_X^{~~0.17\pm 0.10}$ & OLS $\rm Y(X)$ & 0.26 & 0.137540 \\
\hline
\multicolumn{5}{c}{\small{Substructure Level vs. Physical Parameters} } \\
\hline
\end{tabular}
\caption{Table with best fits.  
Temperature is given in keV, 
mass in $10^{14}\rm M_\odot$ and luminosity in $10^{44}\rm erg/s$.
} \label{estatistica}
\end{table*}


\subsection{Clusters Parameters}

In table \ref{clusters_parameters} we give the corrected substructure level ($S_C$), $M_{500}$, X-ray luminosity,
temperature, redshift, the 2D analytical surface
brightness model fitted, core radius ($R_C$) and signal-to-noise of the 34 clusters of the sample.

We computed the substructure level and the 2D analytical surface brightness
model, while the other parameters were obtained from the literature
\citep{2002Reiprich,2004Sun,2007Chen,2011Maughan}. $M_{500}$ and X-ray luminosity
were corrected to a Hubble constant of 70 km s$^{-1}$ Mpc$^{-1}$, and then X-ray luminosity 
was extrapolated to the bolometric band (0.01 - 100 keV) using K-correction and the XSPEC 12.0 MEKAL model
(Mewe-Kaastra-Leidahl plasma emission code), since literature values were given for different Hubble constants and 
energy bands.
Redshifts were obtained from \textsc{NED} (NASA/IPAC Extragalactic Database) with uncertanties varying from $10^{-6}$ to $10^{-4}$,
therefore as they are extremly small compared to other uncertainties we use, 
they are not displayed in Table \ref{clusters_parameters}.

\begin{table*}
\small
\centering
\begin{tabular}{|c|c|c|c|c|c|c|c|c|c|}
\hline 
Cluster & $S_C$ & $M_{500}, 10^{14}\rm M_\odot$ &
$L_X, 10^{44}\rm erg~s^{-1}$ & $kT$, keV & $z$ & Model & R & $R_C$ & $S/N$ \\
\hline





\hline

A85               & $ 0.095^{+0.009}_{-0.007} $ & 5.77 $\pm$ 1.12 & 11.88 $\pm$ 0.30 & $ 6.51^{+0.16}_{-0.23} $ & 0.055 & $\beta$ & C & 9  & 735 \\
A399      	  & $ 0.101^{+0.055}_{-0.031} $ & 5.53 $\pm$ 1.54 & 8.72 $\pm$ 1.07 & $ 6.46^{+0.38}_{-0.36} $ & 0.071 & $\beta$  & C & 76 & 403 \\
A401      	  & $ 0.023^{+0.007}_{-0.008} $ & 5.99 $\pm$ 0.87 & 16.14 $\pm$ 0.65 & $ 7.19^{+0.28}_{-0.24} $ & 0.074 & $\beta$ & C & 55 & 360\\
A478      	  & $ 0.065^{+0.016}_{-0.006} $ & 6.32 $\pm$ 2.14 & 22.16 $\pm$ 0.97 & $ 6.91^{+0.40}_{-0.36} $ & 0.088 & $\beta$  & C & 14 & 246 \\
A520      	  & $ 0.120^{+0.225}_{-0.038} $ & 7.83 $\pm$ 0.32 & 18.40 $\pm$ 0.25 & $ 6.59^{+0.22}_{-0.23} $ & 0.199 & $\beta$ & M & 77 & 195 \\
A644      	  & $ 0.077^{+0.038}_{-0.013} $ & 6.01 $\pm$ 1.54 & 10.28 $\pm$ 0.43 & $ 6.54^{+0.27}_{-0.26} $ & 0.070 & $\beta$ & C & 43 & 418 \\
A665      	  & $ 0.159^{+0.022}_{-0.018} $ & 9.62 $\pm$ 0.45 & 23.40 $\pm$ 0.26 & $ 7.48^{+0.28}_{-0.28} $ & 0.182 & S\'ersic & M & 17 & 175 \\
A907      	  & $ 0.062^{+0.010}_{-0.005} $ & 4.90 $\pm$ 0.23 & 11.50 $\pm$ 0.10 & $ 5.16^{+0.10}_{-0.10} $ & 0.153 & $\beta$ & M & 12 & 218 \\
A1204     	  & $ 0.043^{+0.013}_{-0.016} $ & 2.97 $\pm$ 0.23 & 9.95 $\pm$ 0.19 & $ 3.41^{+0.07}_{-0.12} $ & 0.171 & $\beta$ & M & 5 & 130 \\
A1413     	  & $ 0.052^{+0.005}_{-0.007} $ & 7.11 $\pm$ 0.28 & 17.10 $\pm$ 0.10 & $ 7.21^{+0.16}_{-0.16} $ & 0.143 & $\beta$ & M & 20 & 364\\
A1644     	  & $ 0.152^{+0.027}_{-0.022} $ & 5.24 $\pm$ 3.07 & 4.04 $\pm$ 0.77 & $ 4.70^{+0.90}_{-0.70} $ & 0.047 & $\beta$  & C & 10 & 234\\
A1650     	  & $ 0.034^{+0.008}_{-0.014} $ & 4.66 $\pm$ 1.55 & 8.35 $\pm$ 1.16 & $ 5.68^{+0.30}_{-0.27} $ & 0.084 & $\beta$  & C & 23 & 356 \\
A1689     	  & $ 0.051^{+0.006}_{-0.006} $ & 9.36 $\pm$ 0.59 & 39.40 $\pm$ 0.30 & $ 9.02^{+0.27}_{-0.27} $ & 0.183 & $\beta$  & M & 12 & 392 \\
A1795     	  & $ 0.099^{+0.008}_{-0.010} $ & 7.05 $\pm$ 2.75 & 11.92 $\pm$ 0.36 & $ 6.17^{+0.26}_{-0.25} $ & 0.062 & $\beta$  & C & 16 & 445\\
A1914     	  & $ 0.112^{+0.040}_{-0.012} $ & 9.17 $\pm$ 0.63 & 34.30 $\pm$ 0.34 & $ 9.59^{+0.33}_{-0.33} $ & 0.171 & $\beta$ & M  & 30 & 197\\
A2029     	  & $ 0.021^{+0.006}_{-0.012} $ & 7.11 $\pm$ 2.35 & 23.50 $\pm$ 0.93 & $ 7.93^{+0.39}_{-0.36} $ & 0.078 & S\'ersic & C & 12 & 298 \\
A2034 	  	  & $ 0.092^{+0.117}_{-0.033} $ & 5.87 $\pm$ 0.19 & 9.41 $\pm$ 0.07 & $ 6.59^{+0.15}_{-0.15} $ & 0.113 & $\beta$ & M & 73 & 292 \\
A2142     	  & $ 0.023^{+0.059}_{-0.011} $ & 10.24 $\pm$ 2.60 & 30.13 $\pm$ 1.55 & $ 8.46^{+0.53}_{-0.49} $ & 0.091 & S\'ersic  & C & 12 & 554 \\
A2163     	  & $ 0.155^{+0.022}_{-0.024} $ & 24.32 $\pm$ 1.82 & 93.90 $\pm$ 1.33 & $ 14.70^{+0.87}_{-0.85} $ & 0.203 & $\beta$  & M & 36 & 475 \\
A2319     	  & $ 0.085^{+0.090}_{-0.035} $ & 9.69 $\pm$ 1.54 & 24.05 $\pm$ 0.86 & $ 8.84^{+0.29}_{-0.24} $ & 0.056 & $\beta$  & C & 78 & 436\\
A2657     	  & $ 0.040^{+0.010}_{-0.011} $ & 4.33 $\pm$ 0.94 & 1.56 $\pm$ 0.05 & $ 3.53^{+0.12}_{-0.12} $ & 0.040 & $\beta$  & C & 28 & 243 \\
A3158     	  & $ 0.038^{+0.032}_{-0.012} $ & 4.11 $\pm$ 0.64 & 6.23 $\pm$ 0.33 & $ 5.41^{+0.26}_{-0.24} $ & 0.060 & $\beta$  & C & 59 & 417\\
A3266     	  & $ 0.137^{+0.036}_{-0.034} $ & 13.74 $\pm$ 3.40 & 11.68 $\pm$ 0.46 & $ 7.72^{+0.35}_{-0.28} $ & 0.059 & $\beta$  & C & 58 &479 \\
A3562     	  & $ 0.055^{+0.019}_{-0.015} $ & 2.51 $\pm$ 0.31 & 3.09 $\pm$ 0.13 & $ 4.47^{+0.23}_{-0.21} $ & 0.049 & $\beta$  & C & 30 & 247\\
A3921     	  & $ 0.077^{+0.015}_{-0.019} $ & 4.71 $\pm$ 1.07 & 5.35 $\pm$ 0.36 & $ 5.39^{+0.38}_{-0.35} $ & 0.093 & $\beta$  & C & 35 & 233\\
A4038             & $ 0.004^{+0.026}_{-0.006} $ & 1.54 $\pm$ 0.06 & 1.67 $\pm$ 0.22 & $ 3.15^{+0.03}_{-0.03} $ & 0.030 & $\beta$  & RB & 37 &456 \\
ESO3060170-B $\dagger$ & $ 0.082^{+0.014}_{-0.018} $ & 1.50 $\pm$ 0.50 & 0.65 $\pm$ 0.04 & $ 2.63^{+0.05}_{-0.05} $ & 0.039 & S\'ersic  & S & 13 & 120\\
EXO0422           & $ 0.033^{+0.006}_{-0.005} $ & 1.94 $\pm$ 1.22 & 1.68 $\pm$ 0.40 & $ 2.90^{+0.90}_{-0.60} $ & 0.040 & $\beta$  & C & 9 & 189 \\
MKW 3S            & $ 0.040^{+0.010}_{-0.005} $ & 2.30 $\pm$ 0.66 & 2.48 $\pm$ 0.09 & $ 3.45^{+0.13}_{-0.10} $ & 0.045 & $\beta$  & C & 23 & 528\\
MS 0906.5+1110    & $ 0.112^{+0.015}_{-0.012} $ & 4.53 $\pm$ 0.25 & 9.01 $\pm$ 0.15 & $ 5.19^{+0.17}_{-0.17} $ & 0.180 & S\'ersic  & M & 3 & 110\\
PKS0745-191       & $ 0.032^{+0.005}_{-0.003} $ & 6.34 $\pm$ 0.25 & 35.90 $\pm$ 0.68 & $ 7.21^{+0.11}_{-0.11} $ & 0.103 & $\beta$ &RB & 11 & 251\\
RXCJ 1504-0248    & $ 0.074^{+0.045}_{-0.026} $ & 9.81 $\pm$ 1.13 & 66.60 $\pm$ 0.72 & $ 7.13^{+0.24}_{-0.24} $ & 0.109 & S\'ersic & M  & 1 &187 \\
RXJ1720.1+2638    & $ 0.061^{+0.008}_{-0.009} $ & 6.83 $\pm$ 0.38 & 22.30 $\pm$ 0.23 & $ 5.87^{+0.12}_{-0.12} $ & 0.164 & $\beta$  & M & 7 & 199\\
ZWCL1215          & $ 0.125^{+0.049}_{-0.051} $ & 6.76 $\pm$ 4.10 & 6.27 $\pm$ 1.73 & $ 6.36^{+2.94}_{-2.01} $ & 0.075 & $\beta$  & C & 66 & 193\\
\hline 
\end{tabular}
\caption{Clusters parameters. Core radius ($R_C$) is given in pixel units (whole image is 500 $\times$ 500 pixels).
Literature references (R) are given by: RB - \citet{2002Reiprich}, S - \citet{2004Sun}, 
C - \citet{2007Chen}, M - \citet{2011Maughan}. 
$\dagger$ The mass is given within $1/3 R_{vir}$, $R_{vir} = 1.35$ Mpc.}\label{clusters_parameters}
\end{table*}


\subsection{Substructure Level vs. Physical Parameters}

Keeping in mind the different statistical approaches, for all the correlations
between the substructure level and physical parameters the OLS $(Y\vert X)$ 
was used, since the substructure level may depend on mass, temperature 
and luminosity but these quantities should not be dependent on substructure, the way it was defined.
For the scaling relations we used the OLS (Bisector) since temperature, luminosity 
and mass have complicated relations connecting them. For instance,
mass is one of the quantities which determines the cluster temperature, but temperature is used
to compute the mass. Luminosity is the observed quantity (flux and redshift), although it depends on
temperature. Therefore complicated relations exist between them, which made us use the
OLS (Bisector). On the other hand, concerning the data size, since we have used 34 data points
we chose the Jackknife resampling method to perform all the fits. The results of the fits
are presented in Table \ref{estatistica}. 




\subsubsection{Substructure Level vs. Redshift}

\begin{figure}[!htb]
\centering
\includegraphics[width=0.47\textwidth]{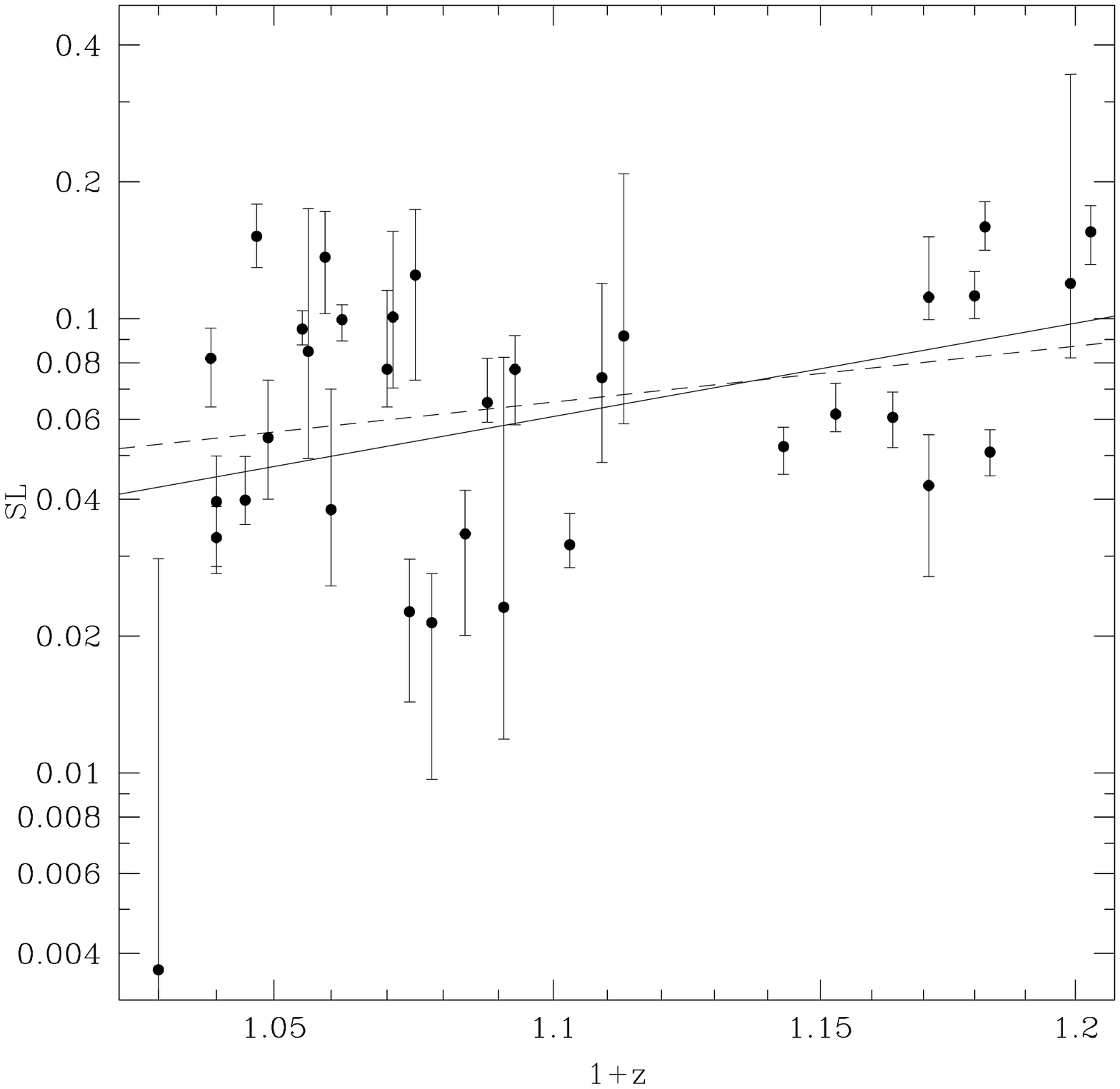}
\caption{\small{Substructure level evolution. Solid and dashed lines correspond to the fit
using all 34 clusters, and excluding Abell 4038 (the cluster with $S_C < 0.004$, on the bottom left
of the plot), respectively. The fit with all clusters is given by 
$SL = 0.036^{+0.010}_{-0.007} \times (1+z)^{5.45 \pm 1.95}$, where we used the OLS $\rm Y(X)$. The Pearson
correlation coeficient obtained was 0.35.}}\label{gsub_z}
\end{figure}

\begin{figure*}[!ht]
\centering
\includegraphics[width=1.\textwidth]{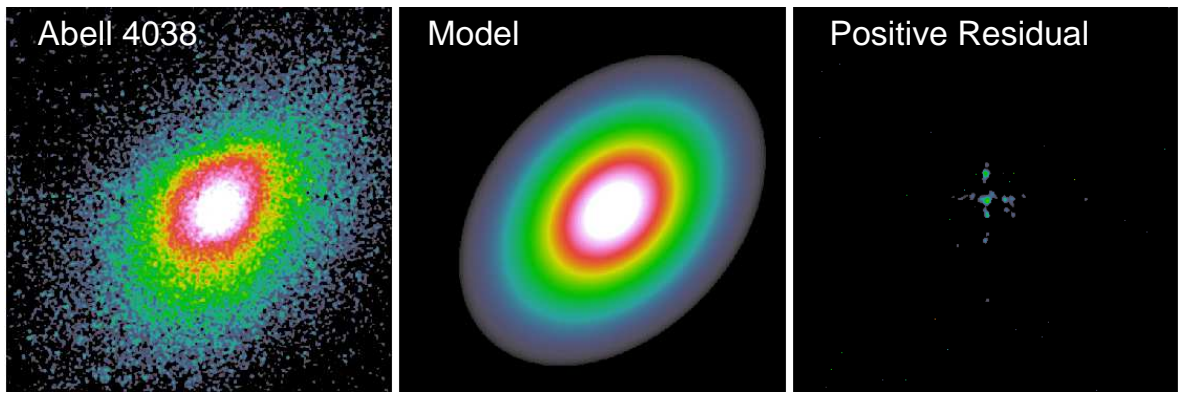}
\caption{\small{Abell 4038, a symmetric cluster, with few substructures. The substructures are basically
due to the ACIS-I chip gaps, which represent only a very small fraction of the total counts ($S_C = 0.004$).}}\label{A4038}
\end{figure*}

\begin{figure*}[!ht]
\centering
\includegraphics[width=1.\textwidth]{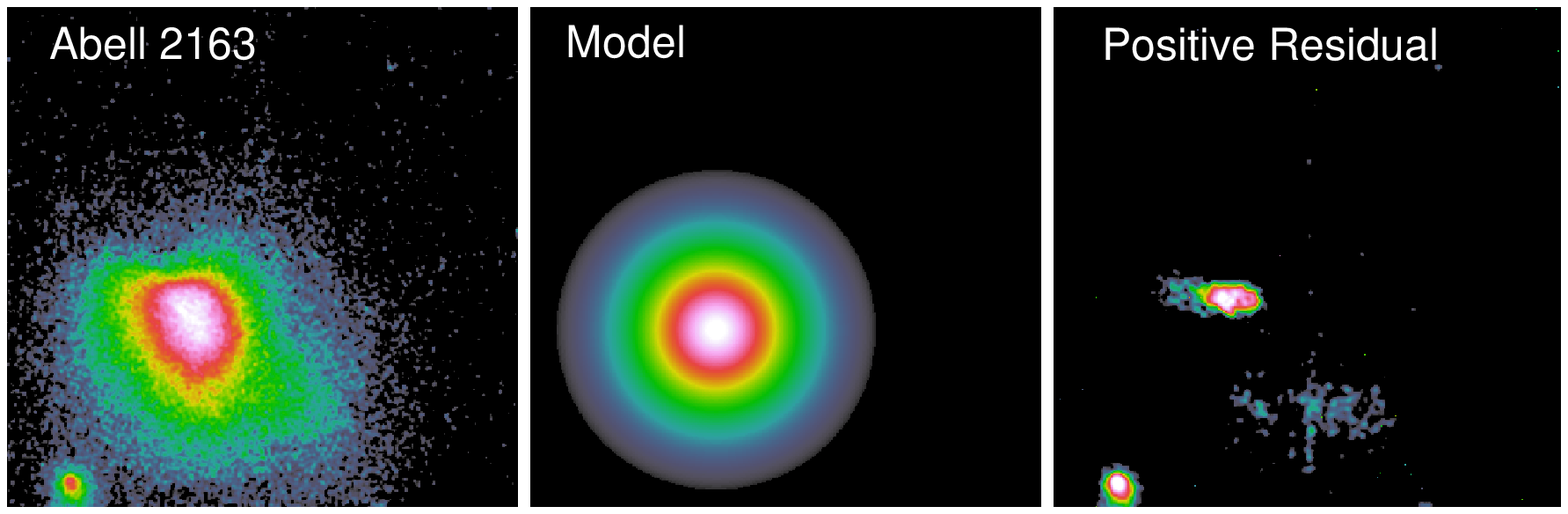}
\caption{\small{Abell 2163, a perturbed cluster, with asymetries and substructures ($S_C = 0.155$).}}\label{A2163}
\end{figure*}

Figure \ref{gsub_z} shows the substructure level as a function of cluster
redshift. 
We see that there is a dependence between the substructure level and
redshift ($S \propto (1+z)^{5.45 \pm 1.95}$), although within $2.8 \sigma$ we 
find no evolution at all in the substructure level. The dependence on redshift
may be explained by the fact that nearby clusters fill a larger detector area compared to
more distance clusters, and as explained in the calibration Section (\S~\ref{calibration_section}),
 they tend to have the 
substructure level understimated, since substructures that lie within small clustercentric 
distances are incorporated into the surface brightness fit and are hardly quantified. 
Furthermore the Pearson correlation coefficient of 0.35 shows us a weak
correlation, which translates as no significant structural evolution of the
gas distribution. 
We also note the strong scatter of the data points in the
redshift range $z \in [0.02,0.2]$, showing that we find clusters in very
different dynamical states in this redshift interval, from those highly
symmetrical to the very disturbed ones (see Figures \ref{A4038} and \ref{A2163}).
Such a scatter may be related to the young (dynamically speaking) age of massive clusters. Abell 4038,
which has the smallest substructure level, has been considered in all fits, however we present in Figures
\ref{gsub_z}, \ref{gsub_T}, \ref{gsub_Lx} and \ref{gsub_Mtot} dashed lines representing the fits excluding it,
since one could ask how much influence it has in determining the slopes of the curves.

Computing a temperature map based on \textit{Chandra} data,
\citet{2001MarkVik} clearly showed that Abell 2163 cluster (see Figure \ref{A2163}) is undergoing a major
merger, which explains its high substructure level ($SL = 0.155^{+0.022}_{-0.024}$).


\subsubsection{Substructure Level vs. Temperature}

\begin{figure}[!htb]
\centering
\includegraphics[width=0.47\textwidth]{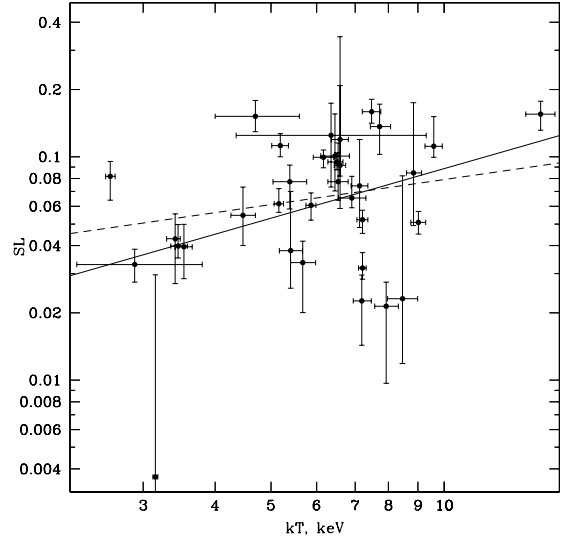}
\caption{\small Substructure Level vs. Temperature. For explanation on the fits, 
see caption on Figure \ref{gsub_z}.
The solid line is given by $SL = 0.016^{+0.012}_{-0.007} \times kT^{0.74 \pm 0.30}$, 
where the fits performed was the OLS $\rm Y(X)$ and the Pearson coefficient 
obtained was 0.36.}\label{gsub_T}
\end{figure}

We see in Figure \ref{gsub_T} that hotter clusters show the tendency to present more
substructures, however the Pearson coefficient of 0.36 shows a weak correlation.
The slope of $0.74 \pm 0.30$ shows a positive corrrelation 
within $2.5 \sigma$, however the weak correlation does not allow us to state any firm
conclusion concerning the intensity of substrutures and gas temperature enhancements. 

\subsubsection{Substructure Level vs. Luminosity}

\begin{figure}[ht!]
\centering
\includegraphics[width=0.47\textwidth]{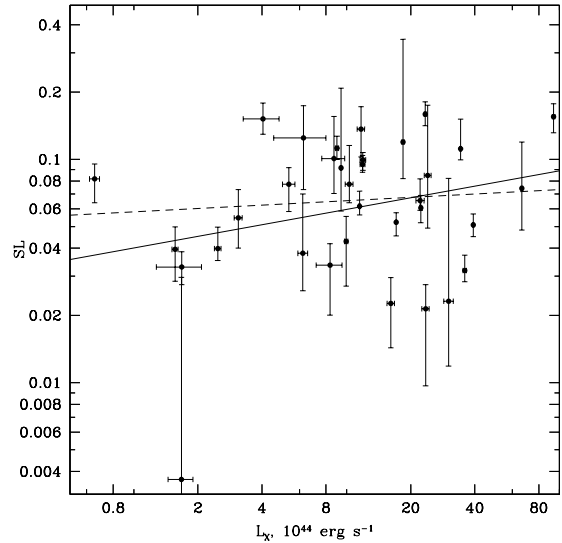}
\caption{\small Substructure Level vs. Luminosity. For explanation on the fits, 
see caption on Figure \ref{gsub_z}. The solid line is given by
$SL = 0.040^{+0.014}_{-0.010} \times L_X^{~~0.17\pm 0.10}$. The fits performed was the OLS $\rm Y(X)$
. The Pearson coefficient obtained was 0.26.}\label{gsub_Lx}
\end{figure}

We see in Figure \ref{gsub_Lx} that substructure is basically independent of X-ray luminosity, 
the Pearson coefficient of 0.26 being the lowest 
between the substructure level and the physical parameters. Furthermore, the substructure
level is compatible with no dependency at all with the X-ray luminosity within only
$1.7 \sigma$.

\subsubsection{Substructure Level vs. Mass}

\begin{figure}[ht!]
\centering
\includegraphics[width=0.47\textwidth]{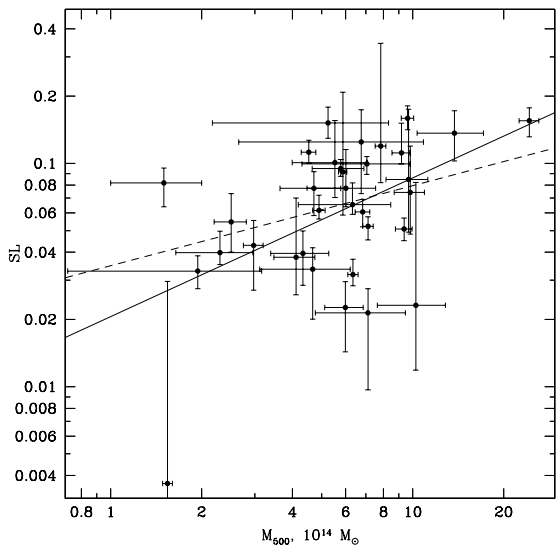}
\caption{\small Substructure level dependency on mass. For explanation on the fits, 
see caption on Figure \ref{gsub_z}. 
The solid line is given by $SL= 0.021^{+0.010}_{-0.007} \times M_{500}^{~~0.62 \pm 0.20}$,
where the fits performed was the OLS $\rm Y(X)$ and the obtained Pearson coefficient was 0.49.}\label{gsub_Mtot}
\end{figure}

Figure \ref{gsub_Mtot} shows the substructure level as a function of the
cluster mass. We see that more massive clusters show the tendency to
present more substructures. The relation between the substructure level and 
the mass has the form $S \propto M^{0.62 \pm 0.20}$, with a positive 
correlation within $3.1 \sigma$, however, different from the temperature, luminosity 
and redshift correlations with the substructure level, the Pearson correlation coeficient
is the largest (0.49) between them, presenting a 
strong (Null Hypothesis Significance= 0.003265 -- there is only $\sim 0.3 \%$ probability 
of not presenting correlation) relation between the amount of 
substructures a cluster presents and its mass.  

\subsection{Scaling Relations}

\begin{figure}[ht!]
\centering \scalebox{0.38}{\includegraphics{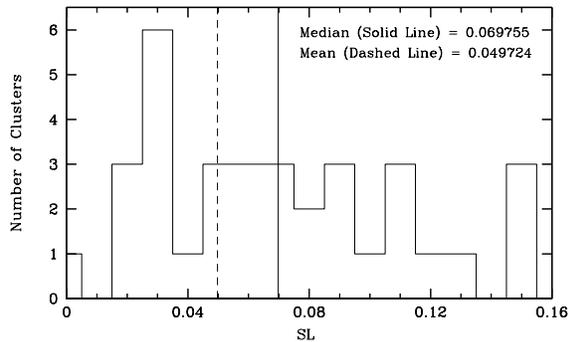}}
\caption{\small Substructure level distribution of clusters.}\label{hist_SL}
\end{figure}

\citet{2007Chen} constructed two cluster samples based on the intra-cluster
plasma central temperature, and they concluded that cooling-core clusters have
different scaling relations compared to non-cooling-core clusters. Using the
same idea, we created two different groups based on the substructure
level. We computed the mean and median substructure level of the sample
(illustrated in Figure \ref{hist_SL}). We choose the median, which is more
robust regarding extreme data points, as a division line between high and low
substructure level. Therefore, clusters were separated into two sub-groups
according to their substructure level compared to the median value of the whole sample.
Numerically, a cluster was considered highly
substructured if its substructure level were greater than $S = 0.069755$, and a
low substructure level cluster otherwise.

\begin{figure}[ht!]
\centering
\includegraphics[width=0.47\textwidth]{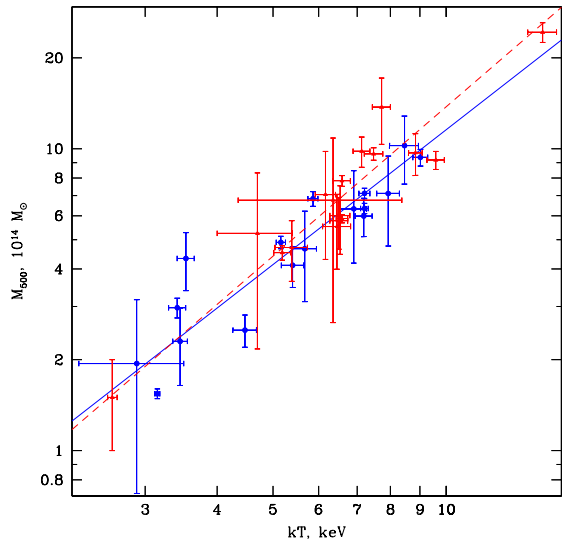}
\caption{\small Mass vs. Temperature. In red (triangles) and blue (circles) 
are the high and low substructure level clusters, respectively. The red (dashed)
and blue (solid) lines are the best fit for the high and low substructure level clusters,
respectively. The fit of the low substructure clusters is given by 
$M_{500}=0.378^{+0.079}_{-0.065}\times kT^{1.49\pm 0.10}$, where the fit performed was the OLS (Bisector)
and the Pearson coefficient obtained was 0.92. On the other hand the high substructure level clusters
fit is given by $M_{500}=0.312^{+0.039}_{-0.036}\times kT^{1.64\pm 0.07}$ with a Pearson coefficient of 0.94.}\label{Mtot_T}
\end{figure}

\begin{figure}[ht!]
\centering
\includegraphics[width=0.47\textwidth]{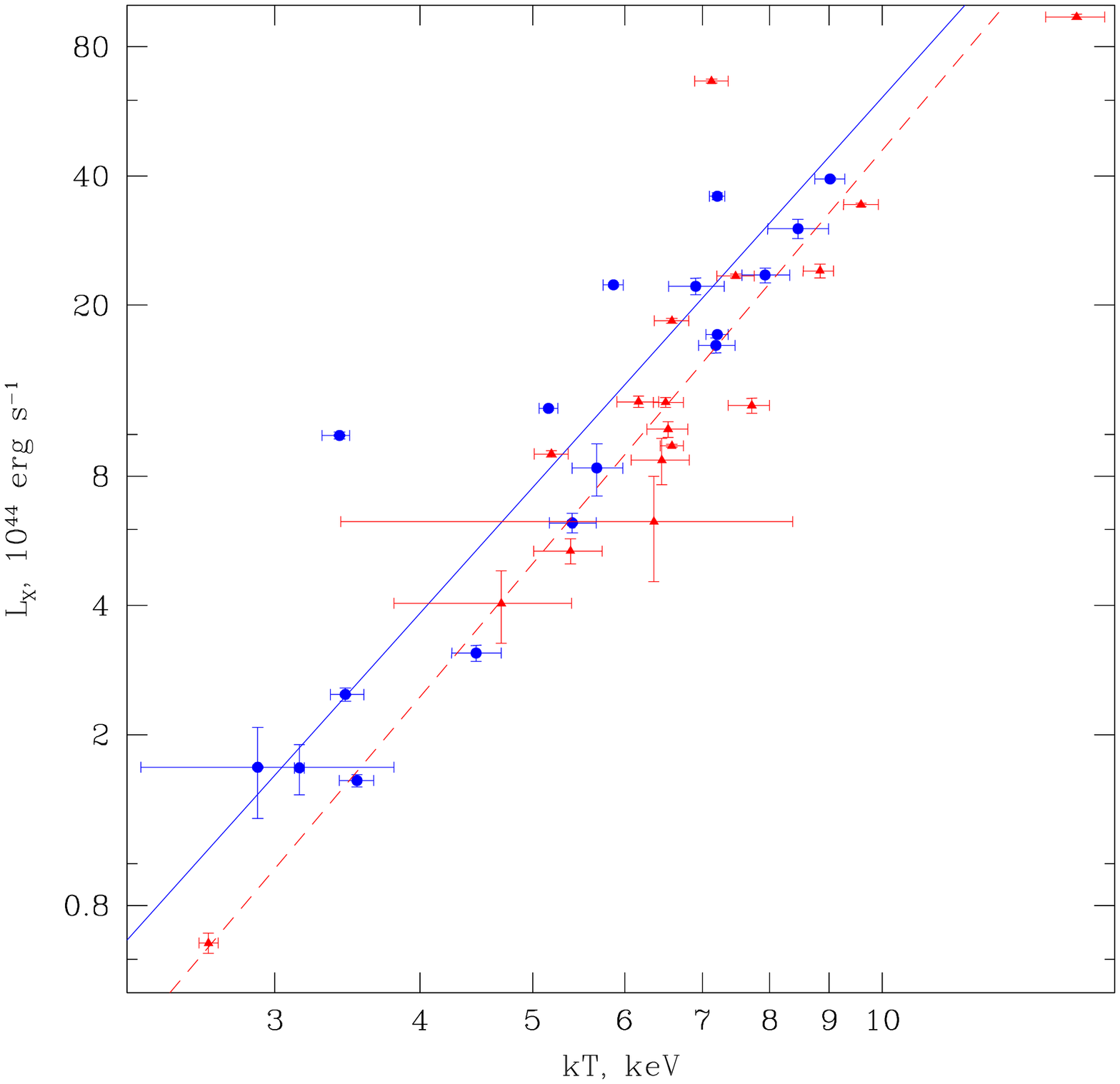}
\caption{\small Luminosity vs. Temperature. See Figure \ref{Mtot_T} for information on the different
points and lines. The fit of the low substructure clusters is given by 
$L_X=0.058^{+0.024}_{-0.017}\times kT^{3.02\pm 0.18}$, where the fit performed was the OLS (Bisector)
and Pearson coefficient obtained was 0.90. On the other hand the high substructure level clusters
fit is given by $L_X=0.029^{+0.013}_{-0.009}\times kT^{3.20\pm 0.29}$ with a Pearson coefficient of 0.91.}\label{Lx_T}
\end{figure}

\begin{figure}[ht!]
\centering
\includegraphics[width=0.47\textwidth]{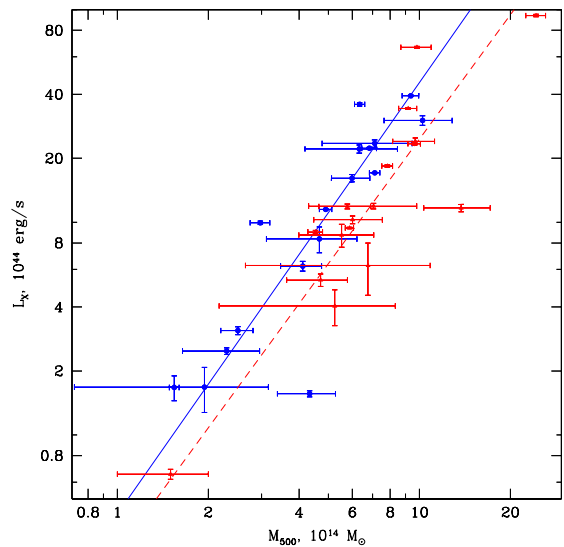}
\caption{\small Luminosity vs. Mass. See Figure \ref{Mtot_T} for information on the different
points and lines. The fit of the low substructure clusters is given by 
$L_X=0.424^{+0.145}_{-0.108}\times M_{500}^{~~2.03\pm 0.15}$ , where the fit performed was the OLS (Bisector)
and Pearson coefficient obtained was 0.88. On the other hand the high substructure level clusters
fit is given by $L_X=0.281^{+0.069}_{-0.055}\times M_{500}^{~~1.94\pm 0.17}$ with a Pearson coefficient of 0.89}\label{Lx_Mtot}
\end{figure}

The cluster segregation in high and low substructure level shows
that hotter clusters are distributed equaly between high and low substructure
level clusters as well as the more massive and luminous clusters do not reside
in a preferential group (high and low substructure groups).

%




The Pearson correlation strength coefficients for the scaling relations of
these different groups are very similar, with the correlations for the
high substructure group being slighly higher (roughly 0.01 above). The slopes
are very similar also, with the excepction of the the $M-T$ relation being 
$1.23 \sigma$ compatible, while the $L-T$ and $L-M$ relations are $0.53 \sigma$ and $0.40 \sigma$ compatibles, respectively.
However, differently from Figure \ref{Mtot_T}
in which the slope is similar and the intercept off-set between the curves is small, Figures \ref{Lx_T} and \ref{Lx_Mtot}
show a clear off-set ($2.07\sigma$ and $2.23\sigma$ for $L-M$ and $L-T$ respectively) 
between the curves for high and low-substructure level clusters. It appears that
given a fixed cluster temperature or mass, the luminosity expected for low-substructure clusters 
tend to be higher. A possible explanation may be that low-substructure clusters, therefore more relaxed ones, have had
enough time for the gas to accomodate into the gravitational potential and become denser,
which enhances the X-ray luminosity. Figure 6 of \citet{2007Chen} shows a very similar effect, where their cool-core clusters
present higher X-ray luminosities, for a fixed temperature, compared to non-cool-core clusters, which would advocate in favor 
of low-substructure clusters, and therefore more relaxed ones being associated with cool-core clusters.

The different scale relations between low-high substructure level clusters suggests that substructures are an important 
factor to bias scaling relations, therefore it  
may affect the mass of clusters determination and thus the mass function that is used to test cosmological models, 
e.g. using the Press-Schechter Extended model etc.


\section{Conclusions}


We have developed a new method to quantify X-ray substructures in clusters of
galaxies based on the ratio between the number of counts in the residual and
original X-ray images. We calibrated the method and  then 
applied it to 34 clusters of galaxies in
order to obtain the substructure level dependence with physical parameters,
such as mass, temperature, X-ray luminosity and redshift.

The calibration was done using Monte Carlo simulations, which showed that 
the method recuperates very well the true amount of substructure for small 
angular core radii clusters (with respect to the whole image size) and good 
signal-to-noise ratio observations.


The substructure level high scatter (spanning from less than 1 to $\simeq 16\%$)
in the redshift range $z \in [0.02,0.2]$
shows that clusters are found in all dynamical states in the local Universe:
from those relaxed to completely disturbed.



We have not found any strong evidence of correlation between the substructure level and 
physical properties of the clusters, gas temperature,
X-ray luminosity and redshift. However, there seems to be a correlation between the
substructure level and the cluster mass, which is given by its Pearson correlation coeficient 
of 0.49. For our sample of 34 clusters it represents a probability of not presenting correlation 
of the order of $0.3\%$.

The distinction between high and low substructure level clusters has shown to
be interesting, since different scaling relations were found with these two
sub-samples (they present an off-set of $\sim 2\sigma$ -- given a fixed mass or temperature,
low substructure clusters tend to be more X-ray luminous), which is an important 
result for cosmological tests which use the 
cluster mass-luminosity relation to compute the mass function. 

A practical application of our method would be the identification of clusters
of very low substructure level. Such relaxed clusters would be ideal
laboratories for studies where the equilibrium hypothesis is of paramount
importance. As an example, the work of Bertolami et al. (2007), on the
interaction between dark matter and dark energy, made use of the Layzer-Irvine
equation, which must hold for a system in virial equilibrium when there is no
interaction in the dark sector. They applied their method to Abell 586,
based on the analysis of Cypriano et al. (2005), which suggests that Abell 586
is indeed a very relaxed cluster.

Finally, it is important to say that the method itself is interesting since
it concerns a new way to quantify substructures in
clusters of galaxies, with a very simple physical interpretation: 
it reflects the fraction of the X-ray luminosity provided by substructures.




\acknowledgments

We are grateful to Ben Maughan, for kindly providing us with new results and Kristian Pedersen, Steen Hansen, Ralph Kraft, William
Forman, Christine Jones, Eric Hallman, Cl\'audia Mendes, 
Hugo Capelato and Ronaldo E. de Souza for interesting discussions.
We are also very grateful to the anonymous referee who helped a lot to improve 
this work.
This work was supported by FAPESP (grants 2008/05970-0 and 2008/04318-7)
and CNPq (\textit{Edital Universal}
472012/2007-0).



{\it Facilities:} \facility{Instituto de Astronomia, Geof\'isica e
Ci\^encias Atmosf\'ericas - Universidade de S\~ao Paulo}.




\begin{thebibliography}{}

\bibitem[Bertolami et al.(2007)]{2007Bertolami} Bertolami, O., Gil 
Pedro, F., \& Le Delliou, M.\ 2007, Physics Letters B, 654, 165 

\bibitem[{{Buote} \& {Tsai}(1995)}]{1995Buote}
{Buote}, D.~A., \& {Tsai}, J.~C. 1995, \apj, 452, 522

\bibitem[{{Buote} \& {Tsai}(1996)}]{1996Buote}
---. 1996, \apj, 458, 27

\bibitem[{{Cavaliere} \& {Fusco-Femiano}(1976)}]{1976Cavaliere}
{Cavaliere}, A., \& {Fusco-Femiano}, R. 1976, \aap, 49, 137

\bibitem[{{Chen} {et~al.}(2007){Chen}, {Reiprich}, {B{\"o}hringer}, {Ikebe}, \&
  {Zhang}}]{2007Chen}
{Chen}, Y., {Reiprich}, T.~H., {B{\"o}hringer}, H., {Ikebe}, Y., \& {Zhang},
  Y.-Y. 2007, \aap, 466, 805

\bibitem[{{Cypriano} {et~al.}(2005){Cypriano}, {Lima Neto}, {Sodr{\'e}},
  {Kneib}, \& {Campusano}}]{2005Cypriano}
{Cypriano}, E.~S., {Lima Neto}, G.~B., {Sodr{\'e}}, Jr., L., {Kneib}, J.-P., \&
  {Campusano}, L.~E. 2005, \apj, 630, 38

\bibitem[{{Demarco} {et~al.}(2003){Demarco}, {Magnard}, {Durret}, \&
  {M{\'a}rquez}}]{2003Demarco}
{Demarco}, R., {Magnard}, F., {Durret}, F., \& {M{\'a}rquez}, I. 2003, \aap,
  407, 437

\bibitem[{{Feigelson} \& {Babu}(1992)}]{1992Feigelson}
{Feigelson}, E.~D., \& {Babu}, G.~J. 1992, \apj, 397, 55

\bibitem[{{Ferrari} {et~al.}(2005){Ferrari}, {Benoist}, {Maurogordato},
  {Cappi}, \& {Slezak}}]{2005Ferrari}
{Ferrari}, C., {Benoist}, C., {Maurogordato}, S., {Cappi}, A., \& {Slezak}, E.
  2005, \aap, 430, 19

\bibitem[{{Henriksen} {et~al.}(2000){Henriksen}, {Donnelly}, \&
  {Davis}}]{2000Henriksen}
{Henriksen}, M., {Donnelly}, R.~H., \& {Davis}, D.~S. 2000, \apj, 529, 692

\bibitem[Hogg et al.(2010)]{2010Hogg} Hogg, D.~W., Bovy, J., 
\& Lang, D.\ 2010, arXiv:1008.4686 

\bibitem[{{Huchra} \& {Geller}(1982)}]{1982Huchra}
{Huchra}, J.~P., \& {Geller}, M.~J. 1982, \apj, 257, 423

\bibitem[{{Isobe} {et~al.}(1990){Isobe}, {Feigelson}, {Akritas}, \&
  {Babu}}]{1990Isobe}
{Isobe}, T., {Feigelson}, E.~D., {Akritas}, M.~G., \& {Babu}, G.~J. 1990, \apj,
  364, 104

\bibitem[{{Jeltema} {et~al.}(2005){Jeltema}, {Canizares}, {Bautz}, \&
  {Buote}}]{2005Jeltema}
{Jeltema}, T.~E., {Canizares}, C.~R., {Bautz}, M.~W., \& {Buote}, D.~A. 2005,
  \apj, 624, 606

\bibitem[{{Jeltema} {et~al.}(2008){Jeltema}, {Hallman}, {Burns}, \&
  {Motl}}]{2008Jeltema}
{Jeltema}, T.~E., {Hallman}, E.~J., {Burns}, J.~O., \& {Motl}, P.~M. 2008,
  \apj, 681, 167

\bibitem[{{Jones} \& {Forman}(1984)}]{1984Jones}
{Jones}, C., \& {Forman}, W. 1984, \apj, 276, 38

\bibitem[{{Jones} \& {Forman}(1992)}]{1992Jones}
{Jones}, C., \& {Forman}, W. 1992, in NATO ASIC Proc. 366: Clusters and
  Superclusters of Galaxies, ed. A.~C. {Fabian}, 49--+

\bibitem[{{Jones} \& {Forman}(1999)}]{1999Jones}
---. 1999, \apj, 511, 65

\bibitem[{{Kauffmann} \& {White}(1993)}]{1993Kauffmann}
{Kauffmann}, G., \& {White}, S.~D.~M. 1993, \mnras, 261, 921

\bibitem[{{Lagan{\'a}} {et~al.}(2008){Lagan{\'a}}, {Lima Neto},
  {Andrade-Santos}, \& {Cypriano}}]{2008Lagana}
{Lagan{\'a}}, T.~F., {Lima Neto}, G.~B., {Andrade-Santos}, F., \& {Cypriano},
  E.~S. 2008, \aap, 485, 633

\bibitem[Lagan{\'a} et al.(2010)]{2010Lagana} Lagan{\'a}, T.~F.,
 Andrade-Santos, F., \& Lima Neto, G.~B.\ 2010, \aap, 511, A15 

\bibitem[{{Lima Neto} {et~al.}(2003){Lima Neto}, {Capelato}, {Sodr{\'e}}, \&
  {Proust}}]{2003LimaNeto}
{Lima Neto}, G.~B., {Capelato}, H.~V., {Sodr{\'e}}, Jr., L., \& {Proust}, D.
  2003, \aap, 398, 31

\bibitem[{{Markevitch} \& {Vikhlinin}(2001)}]{2001MarkVik}
{Markevitch}, M., \& {Vikhlinin}, A. 2001, \apj, 563, 95

\bibitem[Maughan et al.(2011)]{2011Maughan} Maughan, B.~J., Giles, 
P.~A., Randall, S.~W., Jones, C., \& Forman, W.~R.\ 2011, arXiv:1108.1200 

\bibitem[{{Mohr} {et~al.}(1995){Mohr}, {Evrard}, {Fabricant}, \&
  {Geller}}]{1995Mohr}
{Mohr}, J.~J., {Evrard}, A.~E., {Fabricant}, D.~G., \& {Geller}, M.~J. 1995,
  \apj, 447, 8

\bibitem[{{Muanwong} {et~al.}(2006){Muanwong}, {Kay}, \& {Thomas}}]{2006Muan}
{Muanwong}, O., {Kay}, S.~T., \& {Thomas}, P.~A. 2006, \apj, 649, 640

\bibitem[{{Pislar} {et~al.}(1997){Pislar}, {Durret}, {Gerbal}, {Lima Neto}, \&
  {Slezak}}]{1997Pislar}
{Pislar}, V., {Durret}, F., {Gerbal}, D., {Lima Neto}, G.~B., \& {Slezak}, E.
  1997, \aap, 322, 53

\bibitem[Press et al.(1992)]{1992NumRec} Press, W.~H., Teukolsky, 
S.~A., Vetterling, W.~T., 
\& Flannery, B.~P.\ 1992, Cambridge: University Press, |c1992, 2nd ed., 

\bibitem[{{Reiprich} \& {B{\"o}hringer}(2002)}]{2002Reiprich}
{Reiprich}, T.~H., \& {B{\"o}hringer}, H. 2002, \apj, 567, 716

\bibitem[{{Richstone} {et~al.}(1992){Richstone}, {Loeb}, \&
  {Turner}}]{1992Richstone}
{Richstone}, D., {Loeb}, A., \& {Turner}, E.~L. 1992, \apj, 393, 477

\bibitem[{Rodgers \& Nicewander(1988)}]{1988Pearson}
Rodgers, J.~L., \& Nicewander, W.~A. 1988, The American Statistician, 42, 59

\bibitem[{{Sun} {et~al.}(2004){Sun}, {Forman}, {Vikhlinin}, {Hornstrup},
  {Jones}, \& {Murray}}]{2004Sun}
{Sun}, M., {Forman}, W., {Vikhlinin}, A., {Hornstrup}, A., {Jones}, C., \&
  {Murray}, S.~S. 2004, \apj, 612, 805

\bibitem[{{Suwa} {et~al.}(2003){Suwa}, {Habe}, {Yoshikawa}, \&
  {Okamoto}}]{2003Suwa}
{Suwa}, T., {Habe}, A., {Yoshikawa}, K., \& {Okamoto}, T. 2003, \apj, 588, 7

\bibitem[{{Xue} \& {Wu}(2000)}]{2000Xue}
{Xue}, Y.-J., \& {Wu}, X.-P. 2000, \apj, 538, 65

\end{thebibliography}



\appendix

\clearpage



\clearpage









\clearpage




\end{document}